\documentclass[runningheads]{llncs}
\usepackage[T1]{fontenc}
\usepackage{graphicx}
\graphicspath{ {images/} }
\begin{document}
%
\title{The World According to a Social Robot -- Augmenting Human-Robot Dialogue With Vision Language Models}
\titlerunning{Augmenting Human-Robot Dialogue With Vision Language Models}
%
\author{Thomas Sievers\orcidID{0000-0002-8675-0122}}
%
\authorrunning{T. Sievers}
%
\institute{Institute of Information Systems, University of Lübeck, 23562 Lübeck, Germany
\email{t.sievers@uni-luebeck.de}}
\maketitle              
\begin{abstract}
Vision Language Models (VLMs) enable robots to visually perceive their environment as well as the actions and characteristics of their conversation partner or humans in collaboration. Especially for social robots deployed in everyday settings and for uncomplicated, natural use, it is essential that the robot has an understanding of situations that is appropriate to human customs. This paper presents initial experiences with the application of a Mistral AI language model with a Pepper robot for Human-Robot Interaction (HRI) in dialogue, as well as an investigation of the effects of additional visual information on response time in different models. The results show that incorporating visual information adds context to the dialogue with only a moderate increase in response time, enabling both the robot and the human to take into account unspoken elements of the situation. Furthermore, using an LLM hosted in Europe offers a solution that complies with European data protection regulations and can therefore facilitate real-life applications more easily.

\end{abstract}
\section{Introduction}

Robots use many different sensors to perceive their environment, orient themselves within it, and explore it. Most of these sensors function very differently from comparable human sensory organs and give these machines superior abilities. However, when we think about collaboration between humans and robots, or simply a conversation in an informative or social setting, the robot must be able to perceive and refer to objects and situations in a human compatible way. The machine must see the same things as its human counterpart and be capable of reflecting on and discussing them.

With the emergence of Large Language Models (LLMs) and their success in Human-Robot Interaction (HRI) applications, it has become easy and almost normal to communicate with robots using natural language. But language alone sometimes does not convey everything that is important in a dialogue. 
Ambiguities and allusions can often only be resolved when visual information is also provided. Especially for the use of social robots in everyday situations, for example in the home environment or in care, and for uncomplicated, natural use, it is essential that the robot has a level of comprehension that is adequate for human customs.
In general, social robots must be designed and developed in such a way that they can meet the requirements of their social environment and respond appropriately and comprehensibly \cite{Dautenhahn2022}. A human-centered perspective should improve the acceptance of social robots in HRI. The inclusion of visual information helps to correctly classify the context during an interaction. In addition, social norms and expectations should guide the robots' decision-making, and the thoughts, feelings, and intentions of others should be understood in a broader social context.

Vision-Language Models (VLMs) are multimodal AI systems created by combining an LLM with a vision encoder that gives the LLM the ability to `see' \cite{ghosh2024}.
Such systems can be used to convey to a robot, based on visual cues, a natural language description of what is happening in front of it from its own perspective. Furthermore, these observations can be directly utilized by an LLM with additional VLM capabilities for its reasoning on the respective task in order to formulate a response to a human utterance.

This work demonstrates the use of a Mistral AI language model \cite{Mistral} with a Pepper robot and examines the influence of an additional visual component on a dialogical interaction between a human and a robot with regard to the interaction context and possible delays in the robot's response time due to VLM capabilities. Choosing an LLM hosted in Europe offers European users a solution that complies with European data protection regulations.

\section{Related Work}

Robots' understanding of situations in scenarios where they interact with humans is often improved through multimodal reasoning. Integrating language and vision from a robot's perspective through reflection processes helps to go beyond basic navigation and object recognition for robots in environments shared with humans \cite{galatolo2025look}.
This is necessary both for social encounters between humans and robots and for industrial environments involving human-robot collaboration \cite{Kawaharazuka2023,Tan2024,Fan2025}. Abbas et al. presented a system for improving HRI in safety-critical industrial systems by integrating LLMs and VLMs into robot control so that robots can understand natural language commands and perceive their environment \cite{Abbas2024}.

The navigation of robots in dynamic, human-centered environments requires socially acceptable decisions based on a solid understanding of the environment, including spatial-temporal perception and the ability to interpret human intentions \cite{Song2025}. Munje et al. investigated whether VLMs can reliably perform the complex spatiotemporal inferences and intention interpretations required for safe and socially compliant robot navigation, and presented a dataset and benchmark for evaluating VLMs in terms of their scene understanding in real-world social robot navigation scenarios \cite{munje2025socialnavsub}.

To address ethical concerns arising from biases in user data when attempting to create user-specific adaptability in VLMs for HRI, Rahimi et al. developed a framework that integrated multimodal user modeling with bias-aware optimization \cite{rahimi2025}. Asuzu et al. proposed the use of a pre-trained LLM for robot planning, supplemented by a VLM responsible for generating scene descriptions that were integrated into the prompt for contextual grounding \cite{Asuzu2025}.

A tool for improving traditional text-based prompts for LLMs with real-time visual inputs was presented by Abbo et al. \cite{Abbo2025}. Text from dialogues between humans in different environments and a robot were supplemented for prompt engineering by summaries of images from the robot's perspective. 
For visual input, they used images from a video recorded during the conversation. Our implementation uses regularly updated individual images captured by the camera in the robot's head, with the most recent image being integrated into the prompt to the LLM.

\section{Methods}

The Mistral multimodal model Pixtral 12B (pixtral-12b-2409) \cite{agrawal2024pixtral12b} was used for text and image processing during the interaction scenarios.
To compare response times, dialogues with and without images were contrasted. In addition, I also tested the response times using the Open AI model GPT-4o mini \cite{OpenAI}.

The LLM was instructed to relate to the content of an image taken with the robot's camera equipment. The robot's front camera took a picture every four seconds to capture an up-to-date impression of what the robot saw. The interval of four seconds was chosen arbitrarily, but seemed to be a practical compromise between timeliness and resource efficiency.
The latest image was sent to the Application Programming Interface (API) of the model for chat completion together with the person's statement in the system prompt.

\subsection{Humanoid Social Robot Pepper}

To test the capabilities of the selected Mistral AI LLM in a human-robot dialogue, I used the humanoid social robot Pepper. Pepper was developed by Aldebaran and first released in 2015 \cite{Pepper}. The robot is 120 centimeters tall and optimized for HRI. It is able to engage with people through conversation, gestures and its touch screen. The robot features an open and fully programmable platform so that developers can program their own applications using software development kits (SDKs) for programming languages like C++, Python or Java respectively Kotlin \cite{PepperSDK2023}. 

The robot application forwarded the utterances of a human conversation partner, which had been received and processed by Pepper's speech-to-text system, as text input to the LLM API, including the latest image from the robot's front camera. The returned API response was then spoken by the robot. With each API call, the whole previous dialog was transferred to the model, allowing it to constantly `remember' what was previously said and refer to it as the dialog progressed. However, only one image was sent with each API call in order to keep data transfer and model costs low.

\subsection{Comparison of LLM/VLM Mistral AI and OpenAI's GPT}

To get an impression of the influence of using image information in dialogue on possible latencies, I compared the response times of Mistral AI models and \mbox{OpenAI's} GPT-4o mini with and without image interpretation via VLM. Basically, the way the API was used was the same for Mistral and OpenAI, but the models required a slightly different structure in the prompt with regard to visual impressions.
Table~\ref{tab1} and Table~\ref{tab2} show the arithmetic mean of the robot's reaction time in milliseconds for various test runs for 12 to 17 reactions, differentiated according to VLM functionality including image upload and LLM functionality without image integration.

\begin{table}
\centering
\caption{Arithmetic mean of response time for VLM functionality.}
\label{tab1}
\begin{tabular}{|l|l|}
\hline
Pixtral-12b-2409 (Mistral) & GPT-4o mini (OpenAI)\\
\hline
1676 ms & 3203 ms\\
\hline
\end{tabular}
\end{table}

\begin{table}
\centering
\caption{Arithmetic mean of the response time for pure LLM functionality.}
\label{tab2}
\begin{tabular}{|l|l|}
\hline
Mistral-large-latest (Mistral) & GPT-4o mini (OpenAI)\\
\hline
1280 ms & 1582 ms\\
\hline
\end{tabular}
\end{table}

The image size uploaded to the VLM was approximately 300 kilobytes in each case.
The  temperature parameter was set to zero for the LLM in all tests. This parameter controls the diversity of output. Lower values make the model more deterministic, while higher values increase creativity and diversity. The internet connection was also the same for all tests.
The results show that using VLM to incorporate visual data leads to a delay in response time compared to using text-based LLM alone. However, at least for the Mistral AI models used, the difference is only around 400 milliseconds, while the difference for GPT-4o mini is more significant at around one and a half seconds with a generally longer response time.

In general, the complexity of the respective query certainly plays a role. In the experiments, questions of varying complexity were asked, ranging from very simple to specialised topics. However, the questions and topics varied in the different trial runs with the different models. Thus, this test series only provides an indication of the expected response times.

\subsection{Prompting the LLM/VLM}

I used prompts for the system role to instruct the model to execute the tasks as a completion task with zero-shot prompting \cite{Brown20}. The conversation and all prompts were originally in German.
The following system prompt was used to incorporate visual information for the interaction scenarios described in the next chapter.

\textbf{System prompt for VLM:} `You are a robot named Pepper. The image is taken from your perspective and shows you the environment you are in and people you can interact with. You do not describe what can be seen in the image, but rather incorporate the information from the image into your conclusions. If there is a person present, engage in conversation based on your impressions, but do not describe your impressions. Be curious and focus on the person's activity or any accessories present, and keep your comments brief. Include the entire course of the conversation in your analysis of the image. Do not repeat any details of the image that you have already mentioned!'

As Abbo et al. mentioned, it is necessary to instruct the model to keep the answers short and concise and to take into account the environment and possible accessories of the persons \cite{Abbo2025}.

For comparison with the response time without visual information, a significantly simpler prompt was used with the LLM.

\textbf{System prompt for LLM only:} `You are a robot. Your name is Pepper. Keep your comments brief.'

In order to meaningfully incorporate visual information into the dialogue with OpenAI's GPT-4o mini model for testing response times, it was necessary to design the prompt slightly differently than with Mistral's Pixtral-12b-2409. In the OpenAI model, the instruction to take the image content into account had to be placed in the user part of the completion message rather than in the system part. The models differ in their interpretation of the instructions, and experimentation was necessary to achieve the best results.

\section{Interactions}

\begin{figure}
\centering
\includegraphics[width=0.95\textwidth]{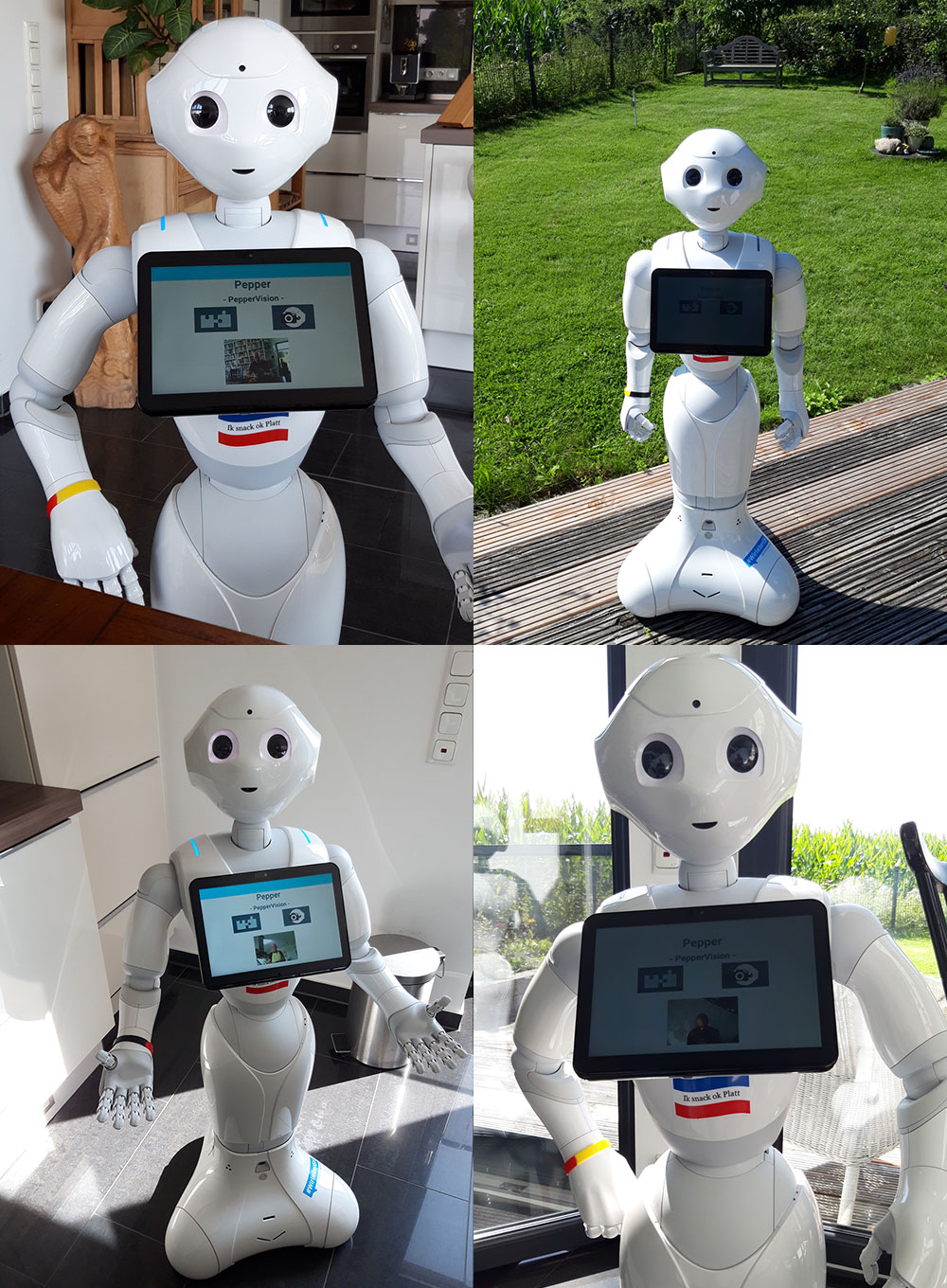}
\caption{The Pepper robot interacting in different example scenarios.}
\label{fig_pepper}
\end{figure}

To test the effectiveness of the proposed use of the Mistral LLM with vision capabilities for interacting with a robot in dialogue, Pepper was positioned in five different scenarios. In addition to the different environments, the human interaction partner wore clothing and carried utensils appropriate to the context, which the robot could take into account in its remarks.
Figure~\ref{fig_pepper} shows the robot placed in example scenarios for interaction with a human. The robot's tablet displayed a photo of the current camera view from the robot's perspective. This allowed the robot's field of vision to be tracked for this experiment.

\textbf{Scenario 1} in a living room with many books on the shelf in the background. The robot Pepper correctly recognized a large bookshelf in the background and a person sitting across from him at the table. The person was drinking from a cup. Pepper asked the person if they liked to read and whether they drank tea or coffee. As the conversation continued, the robot concluded that the person was probably taking a break from reading.

\textbf{Scenario 2} outside on the terrace overlooking the garden. The robot referred to the beautiful weather and the lush greenery of the garden and adjacent fields. When asked specifically about objects, it mentioned a bench that was actually located at the back of the garden.

\textbf{Scenario 3} in the living room with a person sitting on the couch. Pepper commented on the comfort of the couch and referred to the correctly recognized inscription on the person's T-shirt, which raised questions about its meaning (in this case, British television series).

\textbf{Scenario 4} in the kitchen with one person at the stove. The robot recognized the kitchen environment, but due to its angle of view relative to the head of the interaction partner, it was unable to recognize what was on the stove. When a cooking pot was placed in the robot's field of vision, it was correctly recognized and mentioned in the dialogue.

\textbf{Scenario 5} with a person working at a computer. Pepper correctly recognized a work situation at a computer with a monitor, whereby, depending on the viewing angle, with occasionally incomplete coverage of the entire scenario, sometimes a tablet and sometimes the actual laptop was recognized. The robot asked about the type of work being done on the computer and offered assistance.

\section{Discussion}

The inclusion of visual information from the robot's perspective in dialogue with humans enriched the naturalness of the interaction by incorporating context and allowing the robot to proactively refer to the circumstances. Situational elements such as the appearance of the environment, the person's clothing, or objects used were incorporated into the generation of an LLM utterance.
From my own experience, I can say that when the robot proactively references elements of my appearance (e.g. clothing) or my actions, it adds a noticeable new dimension to the interaction, bringing the machine a step closer to the human world.

\section{Limitations}

The language model sometimes tended to describe the entire scene during the course of the dialogue, even when only one detail was actually asked for. Further testing and optimization of prompt engineering could remedy this situation. Occasionally, there were delays of up to three seconds in the generation of the response by the LLM, which would have been slightly shorter without visual components, as a comparison of response times showed. In general, a fluent conversation was possible.

In this experiment, the robot Pepper captured its surroundings with the camera on its forehead between its eyes. Since Pepper also tried to keep its gaze fixed on its conversation partner at all times, the environment outside this field of vision was hardly noticed and thus escaped inclusion in the dialogue. It would be feasible for the robot to be able to avert or lower its gaze, if possible upon request, in order to perceive more details that could be relevant to the conversation.

\section{Conclusion and Future Work}

Particularly for use within the European Union, the ability to use powerful LLMs like Mistral AI from European servers via a simple API integration offers the advantage of not conflicting with European data protection guidelines, as is usually the case when using OpenAI models, for example. This considerably facilitates use in many public institutions such as care facilities or schools.

An extension of the investigations with regard to group interactions or indications from the body language of the interaction partner would be interesting for future work. One could also consider the effects of a type of memory and learning mechanism on the robot's understanding of visual cues. A humanoid robot such as Pepper could possibly use its arms and hands to point at or grasp objects if a correlation could be established between recognized objects in the robot's line of sight and the orientation of its limbs.

In any case, incorporating visual information describing the scene into a robot's dialogue output opens up possibilities for more natural, context-sensitive interaction that is more appropriate for humans.

\begin{credits}
\subsubsection{\discintname}
The author has no competing interests to declare that are relevant to the content of this article.
\end{credits}
%
%
%
\bibliographystyle{splncs04}
\bibliography{references}
\end{document}